\documentclass[preprint,12pt]{elsarticle}

\begin{document}

\begin{frontmatter}

\title{Strain effects on electronic structure and superconductivity in the iron telluride.}

\author[INT]{M. J. Winiarski}
\author[INT]{M. Samsel-Czeka\l a}
\author[IF]{A. Ciechan}
\address[INT]{Institute of Low Temperature and Structure Research, Polish Academy of Sciences, Ok\'olna 2, 50-422 Wroc\l aw, Poland}
\address[IF]{Institute of Physics, Polish Academy of Sciences, al. Lotnik\'ow 32/46, 02-668 Warsaw, Poland}

\abstract{The influence of tensile strain in the {\it ab}-plane on crystal and electronic structure of FeTe has been studied {\it ab initio}. In superconducting FeSe the Fermi surface nesting with a vector $\mathbf{q}\sim(0.5,0.5)\times(2\pi /a)$ is believed to be crucial for rising superconductivity mediated by spin-fluctuations. The results presented here indicate that tensile-strained FeTe also exhibits such conditions. Furthermore, the Fermi surface changes, related to the increase of the lattice parameter {\it a} of this telluride, are opposite to analogous effects reported for FeSe. 
Since a recently reported transition from the double-stripe to the single-stripe magnetic order in FeTe under tensile strain in the {\it ab}-plane is associated with an occurrence of superconductivity in corresponding thin films, these findings allow for drawing a consistent picture of superconductivity in FeSe$_{1-x}$Te$_{x}$ systems, in general.}}

\journal{Intermetallics}

\end{frontmatter}

\section{Introduction}

Iron chalcogenide superconductors draw a wide interest because of their promising applications as well as a possibility of further studies on their complex electronic structure and still unclear mechanism of superconductivity (SC) \cite{Dagotto}. Among them the most often investigated compounds are selenides, whereas the iron telluride exhibits relatively weaker SC but occurring only in its tensile-strained thin films below transition temperature values T$_c$ $<$ 13 K \cite{Han} as well as in some alloys with sulphur below $T_c$ $<$ 10 K \cite{Mizuguchi}. Interestingly, the strain effects on SC in FeSe and FeTe are opposite. In FeSe tensile strain strongly decreases $T_c$ \cite{Nie} but all kinds of compressive strain lead to an increase of its $T_c$ \cite{Nie, Huang, Mizuguchi-FeSe, Margadonna, Millican, Kumar, Okabe, Bellingeri, Bellingeri2, Si}.
The influence of strain on SC of iron chalcogenides is somewhat related to the changes of tetrahedral coordination of Fe atoms \cite{Okabe, Huang, Bellingeri2}. Furthermore, a magnetic order of particular compounds is also  connected with chalcogen atom position in the unit cell (u.c.) of PbO-type \cite{Moon,Ciechan}.

The electronic structure of FeTe has been investigated theoretically \cite{Ciechan,Subedi,Ma,Kumar2,Li} and reveals similar properties to that of FeSe. The density of states (DOS) at the Fermi level ($E_F$) in these iron-based compounds is dominated by the Fe 3d orbital contributions. The Fermi surface (FS) of the 11-type iron chalcogenides is formed by holelike ($\beta$) and electronlike ($\delta$) cylinders around the $\Gamma$ and $M$ points of the Brillouin zone (BZ), respectively. However, the FS nesting vector $\mathbf{q}\sim(0.5,0.5)\times(2\pi /a)$ spanning $\beta$ and $\delta$ FS sheets, postulated to be crucial for spin-fluctuation (SF) mediated SC scenario in iron chalcogenides, in FeTe is not present at all. Recent works have also considered the phononic properties of FeTe \cite{Li,Xia,Okazaki,Gnezdilov} but the SC phenomenon in iron-based chalcogenides is generally believed to be unconventional.

Our former studies on influence of various kinds of strain on electronic structure of FeSe \cite{EPL} and FeSe$_{0.5}$Te$_{0.5}$ \cite{JAC} showed relations between the FS-nesting modifications and possibly SF-mediated SC properties in iron chalcogenides. In these compounds compressive strain diminishes intensity of nesting vector $\mathbf{q}\sim(0.5,0.5)\times(2\pi /a)$, while the tensile strain in the {\it ab}-plane completely destroys these subtle features. The modifications of electronic structure correspond to changes of T$_c$ in specifically strained iron chalcogenides. 

In this work, electronic and structural properties of tensile-strained FeTe are considered in a similar manner to that for FeSe. A qualitative and quantitative description of the intensity of the nesting vector $\mathbf{q}\sim(\pi,\pi)$ for this compound is presented with a discussion of common properties of the electronic structure of iron chalcogenides. The main aim of this study is a simulation of the electronic structure of superconducting FeTe thin films with a particular focus on its relations to unstrained bulk FeTe not exhibiting SC.

\section{Computational details}

Band structure calculations for FeTe have been carried out in the framework of the density functional theory (DFT). A full optimization of the atomic positions and geometry of the tetragonal u.c. of the PbO-type ($P4nmm$, No. 129) under tensile strain in the {\it ab}-plane was performed with the Abinit package \cite{Abinit, PAW}, using Projector Augmented Wave (PAW) pseudopotentials, generated with Atompaw software \cite{Atompaw}. The local density approximation (LDA) \cite{JP} was employed. The 3s3p3d;4s4p states for Fe atoms as well as the 4s4p4d;5s5p states for the Te atoms were selected as a valence-band basis.
Based on these results, the full potential local-orbital (FPLO) band structure code \cite{FPLO} was used in the scalar-relativistic mode to compute the DOSs and Fermi surfaces. Since the FS nesting features  of the 11-type compounds are tiny, very dense {\bf k}-point meshes in the BZ had to be used, {\it i.e.} 64$\times$64$\times$64 and 256$\times$256$\times$256 for the self-consistent field (SCF) cycle and FS maps, respectively.

Finally, a nesting function was determined numerically by the formula: 

\begin{equation}
f_{nest}(\mathbf{q})=\Sigma_{\mathbf{k},n,n'}
\frac{[1-F_{n}^{\beta}({\mathbf{k}})]F_{n'}^{\delta}(\mathbf{k+q})}
{|E_{n}^{\beta}({\mathbf{k}})-E_{n'}^{\delta}(\mathbf{k+q})|},
\end{equation}

where $F_{n}^{\beta}$ and $F_{n'}^{\delta}$ are the Fermi-Dirac functions of states $n$ and $n'$ in bands $\beta$ and $\delta$, ($F=$ 0 or 1 for holes or electrons), respectively. $E_{n}^{\beta}$ and $E_{n'}^{\delta}$ are energy eigenvalues of these bands. The studied $f_{nest}({\mathbf{q||Q}})$, were $\mathbf{Q}=(0.5,0.5)\times(2\pi /a)$ is the ideal nesting vector, represents a frequency of an occurrence of a given vector $\mathbf{q}\sim(\pi,\pi)$ (having its length close or equal to that of $\mathbf{Q}$) in the {\bf k}-space, spanning the FS sheets originating from the $\beta$ and $\delta$ bands.

\section{Results and discussion}

The optimized (LDA) values of the lattice parameters {\it a} and {\it c} as well as the Te anion height in u.c. of unstrained FeTe are collected in Table \ref{table1} Oppositely to our former investigations for FeSe \cite{EPL}, the LDA results of the equilibrium  lattice parameter {\it c} of FeTe are overestimated even in comparison with the GGA data. A similar but less significant effect was also revealed for FeSe$_{0.5}$Te$_{0.5}$ \cite{JAC}. It is worth noting that the presented here results have been obtained with the same PAW pseudoatomic sets, thus the LDA description of crystal structures of iron chalcogenides is insufficient.
Since this issue has been well described by Ricci and Profeta \cite{Ricci}, in this work the discussion of discrepancies between the calculated and experimental values of FeTe lattice parameters is reduced. Due to the fact that the studied here subtle FS changes are closely related to geometry of the u.c. of this compound, the experimental lattice parameter {\it c} \cite{Martinelli} was used for further electronic structure investigations. Only the internal parameter of Te atom position ($h_{Te}$) was optimized for various values of the lattice parameter {\it a}.

\begin{figure}
\includegraphics[scale=1.0]{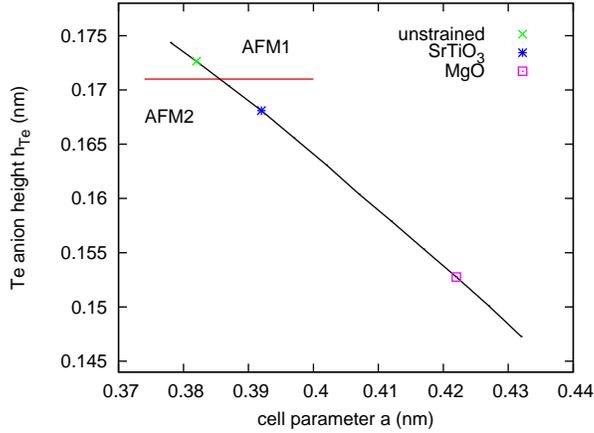}
\caption{Calculated Te anion height ($h_{Te}$) of FeTe for experimental {\it c} \cite{Martinelli} and various values of the lattice parameter {\it a} (solid black line). The results corresponding to bulk FeTe \cite{Martinelli} and SrTiO$_3$ and MgO grown thin films \cite{Han} are marked with symbols. Solid red line denotes transition from AFM1 (double-stripe) to AFM2 (single-stripe) order according to refs. \cite{Moon,Ciechan}.}
\label{Fig1}
\end{figure}

The calculated dependence of $h_{Te}$ in the considered tensile-strained u.c. of FeTe is approximately a linear function of lattice parameter {\it a}, as depicted in Fig. \ref{Fig1}. It is not clear if FeTe thin films exactly adopt lattice parameters of SrTiO$_3$ and MgO \cite{Han}. However, according to refs. \cite{Moon,Ciechan}, even small changes of Fe atom coordination in iron chalcogenides may lead to a transition from the double-stripe (AFM1) to the single-stripe (AFM2) magnetic order. Since in superconducting FeSe the AFM2 order and specific FS nesting features are belived to be crucial for SF-driven pairing mechanism, this transition may also lead to explanation of rising superconductivity in strained FeTe thin films. The AMF2 phase should occur for FeTe samples with increasing the lattice parameter {\it a}, as has been marked by the red solid line in fig. \ref{Fig1}.

\begin{figure}
\includegraphics[scale=1.0]{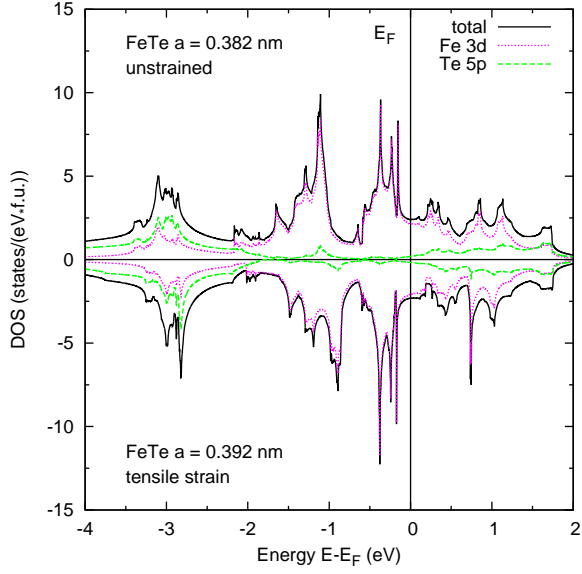}
\caption{Calculated
 DOS of FeTe for experimental lattice parameters, {\it a} = 0.382 nm (unstrained), and tensile-strained, {\it a} = 0.392 nm.}
\label{Fig2}
\end{figure}

The density of states (DOS) at E$_F$, N(E$_F$) = 2.59 states/eV/f.u., for fully optimized (LDA) u.c. of FeTe is significantly higher than N(E$_F$) = 1.43 states/eV/f.u. in FeSe \cite{EPL} and N(E$_F$) = 2.00 states/eV/f.u. in FeSe$_{0.5}$Te$_{0.5}$ \cite{JAC}. The LDA results suggests a linear dependence of N(E$_F$) as a function of the Te content in FeSe$_{1-x}$Te$_{x}$ alloys. The SC properties of iron chalcogenides may be somewhat related to this effect, however the value of $T_c$ $<$ 13 K \cite{Han} in tensile-strained FeTe thin films is lower than $T_c$ = 15 K in bulk FeSe$_{0.5}$Te$_{0.5}$. The values of N(E$_F$) = 2.39 and 2.30 states/eV/f.u.,  calculated for experimental ({\it a} = 0.382 nm) and tensile-strained ({\it a} = 0.392 nm) lattice parameters, respectively, differ only insignificantly.
Furthermore, the overall shapes of FeTe DOS plots, presented in Fig. \ref{Fig2}, are also almost the same for both these conditions. Thus, the strain-induced changes of N(E$_F$) in iron chalcogenides seem to be irrelevant for the SC phenomenon. Similar conclusions have also been drawn for FeSe \cite{EPL} and FeSe$_{0.5}$Te$_{0.5}$ \cite{JAC}.

\begin{figure}
\includegraphics[scale=1.0]{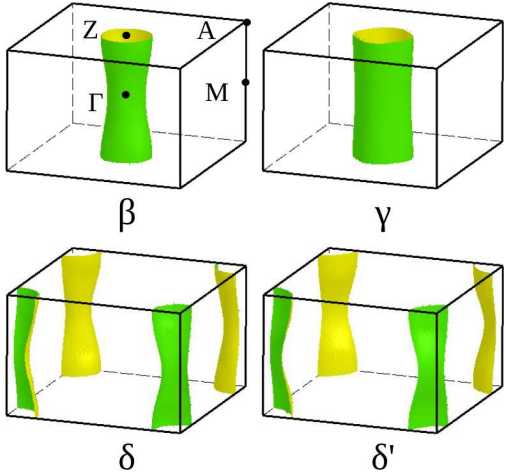}
\caption{Calculated Fermi surface sheets of FeTe for tensile-strained lattice parameter {\it a} = 0.4 nm and bulk value of {\it c}. $\beta$ and $\gamma$ sheets exhibit holelike whereas $\delta$ and $\delta'$ sheets electronlike character.}
\label{Fig3}
\end{figure}

The Fermi surfaces of FeSe$_{1-x}$Te$_{x}$ systems may consist of three holelike sheets ($\alpha$, $\beta$, $\gamma$) around the $\Gamma$ point of the BZ and two electronlike sheets ($\delta$, $\delta'$) around the $M$ point \cite{Subedi, Ciechan_Intermet}. The $\alpha$ sheets in FeSe$_{0.5}$Te$_{0.5}$ \cite{Ciechan_Intermet} and FeTe are partially reduced by chemical pressure, introduced by a substitution of Se with Te atoms that shifts the corresponding band above E$_F$. The influence of tensile strain in the {\it ab}-plane on the topology of FS in FeTe is analogous and leads to the complete lack of the $\alpha$ sheet, as presented in Fig. \ref{Fig3}. However, the other four sheets remain almost unchanged, exhibiting a cylindrical shape, being common in all FeSe$_{1-x}$Te$_{x}$ solid solutions.

\begin{figure}
\includegraphics[scale=1.0]{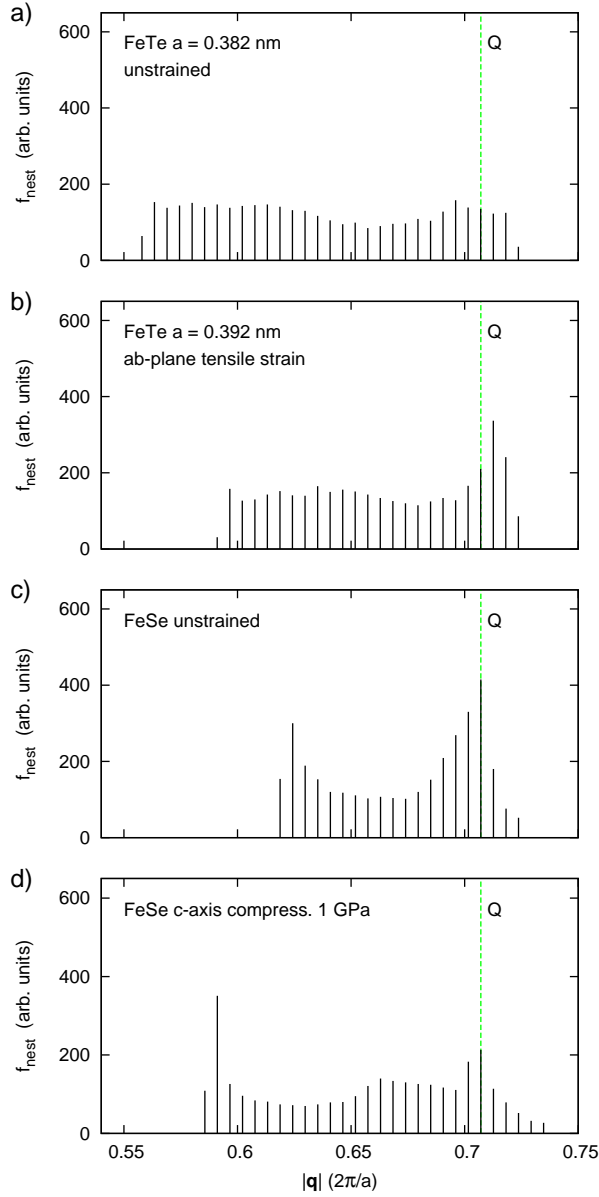}
\caption{Histograms representing the nesting function, $f_{nest}$ vs. lengths of possible vectors $\mathbf{q}$  spanning $\beta$ and $\delta$ FS sheets of FeTe for equilibrium a) and {\it ab}-plane tensile-strained b) lattice parameters, compared to unstrained c) and {\it c}-axis compressive strained FeSe d), taken from ref. \cite{EPL}.}
\label{Fig4}
\end{figure}

A careful analysis of nesting function, $f_{nest}$, between $\beta$ and $\delta$ sheets of unstrained FeTe, depicted in Fig. \ref{Fig4} a), reveals very low intensities of nesting vectors $\mathbf{q}\sim(\pi,\pi)$. However, the overall shape of $f_{nest}$ in this case is similar to that of FeSe under {\it c}-axis compressive strain, seen in Fig. \ref{Fig4} d). It is worth noting that FS nesting properties of tensile-strained FeTe, presented in Fig. \ref{Fig4} b), are very close to those of unstrained FeSe, displayed in Fig. \ref{Fig4} c), thus the influence of strain on the FS of these iron chalcogenides is opposite. The coordination of Fe atoms in FeTe is strongly affected by chemical pressure introduced by Te atoms. Therefore, the intensity of $\mathbf{q}\sim(\pi.\pi)$ in FeTe is diminished, analogously to that of FeSe under some hydrostatic pressure.
These observations may explain the opposite tensile strain effects on nesting-driven, SF-mediated SC in FeSe and FeTe. Furthermore, one can expect that a further strain-induced increase of $T_c$ in FeTe is impossible because of these specific nesting conditions, while the $\mathbf{q}\sim(\pi,\pi)$ is growing with increasing strain.

In unstrained FeTe the double-stripe magnetic order excludes the posibility of SF-mediated superconducting pairing, thus the SC occurs only in tensile-strained samples, in which the single-stripe order should be dominating. The FS nesting properties of tensile-strained FeTe, presented here, suggest that the occurence of SC in this compound is driven by the same mechanism to that in FeSe-based chalcogenides.

\section{Conclusions}

The electronic structure modifications in tensile-strained FeTe thin films are very similar to those of strained FeSe and FeSe$_{0.5}$Te$_{0.5}$. On the one hand, the N(E$_F$) is stable under the strain and the overall changes of electronic structure are rather subtle in these conditions. On the other hand, the strain-induced modifications of the Fermi surface nesting are clearly visible. Since the tensile strain should lead to transition from the double-stripe to the single-stripe magnetic order that favours the FS nesting driven, spin-fluctuation mediated pairing mechanism of superconductivity, presented here results support this idea for all 11-type iron chalcogenides. One can also expect that a further increase of $T_c$ in FeTe is limited by opposite to revealed in FeSe strain-induced changes of intensity of FS nesting $\mathbf{q}\sim(\pi,\pi)$. 

\section*{Acknowledgements}
This work has been supported by the EC through the FunDMS Advanced Grant of the European Research Council (FP7 Ideas) as well as by the National Center for Science in Poland (Grant Nos. 2011/01/B/ST3/02374 and 2012/05/B/ST3/03095). The calculations were partially performed on the ICM supercomputers of Warsaw University (Grant No. G46-13) and in Wroc\l aw Centre for Networking and Supercomputing (Project No. 158).

\begin{table}
\caption{Calculated lattice parameters {\it a}, {\it c}, and free atomic position, $h_{Te}$, in fully optimized u.c. of FeTe in nonmagnetic phase.}
\label{table1}
\begin{tabular}{lllll}
reference &  {\it a} (nm) & {\it c} (nm) & $h_{Te}$ (nm) \\ \hline
this work (LDA) & 0.375 & 0.668 & 0.178 \\
ref. \cite{Ricci} (GGA)  & 0.381 & 0.652 & 0.159 \\
ref. \cite{Martinelli} (expt.) & 0.382 & 0.629 & 0.175 \\
\end{tabular}
\end{table}

\end{document}